\documentclass[aps,pre,twocolumn]{revtex4}%
\usepackage{amsmath, graphicx, subfigure, hyperref}

\begin{document}

\title{Depletion theory and the precipitation of protein by polymer}

\author{Theo Odijk*}

\affiliation{Complex Fluids Theory, Kluyver Laboratory of Biotechnology, Delft University of Technology, Julianalaan 67, 2628 BC Delft, the Netherlands.}

\begin{abstract}

The depletion theory of nanoparticles immersed in a semidilute
polymer solution is reinterpreted in terms of depleted chains of
polymer segments. Limitations and extensions of mean-field theory
are discussed. An explicit expression for the interaction between
two small spheres is derived. The depletion free energy for a particle
of general shape is given in terms of the capacitance or effective
Stokes radius. This affords a close to quantitative explanation for
the effect of polymer on protein precipitation.

\vspace{20 pt}

*Electronic address: odijktcf@wanadoo.nl

\end{abstract}

\maketitle

\begin{table*}[t]
\centering
\begin{tabular}{|c|c|c|c|c|c|c|c|c|c|}
\hline
$R/2a$               & 1     & 1.1   & 1.2   & 1.3   & 1.5   & 2     & 4     & 6      & 10     \\
$2-C_2/(4 \pi a)$ & 0.614 & 0.585 & 0.558 & 0.532 & 0.486 & 0.395 & 0.222 & 0.1538 & 0.0952 \\
\hline
\end{tabular}
\caption{Scaled depletion energy between two nanospheres as a function of their separation}
\end{table*}

It is difficult to set up theories of the polymer distribution near
surfaces because the computed polymer inhomogeneity is sensitive
to the nature of the approximations introduced. Pierre-Gilles de
Gennes devoted considerable time and effort to trying to understand
these problems starting with his survey \cite{1} of forty years ago, which is
still illuminating to read, and culminating in papers containing now
classic ideas like self-similarity \cite{2} and the proximal exponent \cite{3}.
His concise, powerful note \cite{4} on polymer depletion by a small sphere was
strangely neglected by the colloid community for a long time until
the peculiarity of depletion on nanoscales was reassessed merely
a decade ago \cite{5,6}. There has been a flurry of activity in the statistical
physics of nanocolloids immersed in polymer solutions in what has
been termed the protein limit (see e.g. \cite{7,8,9,10,11} and references therein).
Here, I would like to emphasize simple aspects of polymer depletion
by nanoparticles in the spirit of ref. \cite{12}.

Let us recall the argumentation introduced by de Gennes \cite{4} to compute
the free energy of depletion involved in immersing a nanosphere into
a semidilute polymer solution. The solvent is not just ``good''
but needs to be really ``excellent'' (see below), i.e. the excluded volume
$\beta$ between the Kuhn segments equals $A^3$ where $A$ is the segment length.
If the radius $a$ of the sphere is larger than $A$, it is plausible to assume
that $a$ and the polymer correlation length $\xi$ are the only relevant
length scales in the problem. The latter is given by \cite{13}
\begin{equation}
\xi = A^{-5/4} \, c_0^{-3/4} \,,
\end{equation}
where the concentration $c_0$ is the number of polymer segments
per unit volume. For a nanosphere dissolved in a semidilute
solution of low concentration, one readily has $a\!\ll\!\xi$. De Gennes
then argues that there is a volume of order $a^3$ -- independent of $\xi$ --
surrounding the sphere from which polymer is depleted \cite{4}. Hence, the number
of segments depleted is of the order of $a^3 \, c_0$. Since the free
energy of depletion $F_1$ must be proportional to this number, i.e. must
be proportional to the concentration $c_0$, one concludes that \cite{4}
\begin{equation}
F_1 \simeq \left( \frac{a}{\xi} \right)^{\!4/3} \, k_{\rm B} T \,.
\end{equation}
Here, $k_{\rm B}$ is Boltzmann's constant and $T$ is the temperature.

Here, I invoke a different type of scaling argument to derive
Eq.(2) because this will allow a direct assessment of the other aspects
of depletion in the analysis below. The number of polymer segments
depleted from the vicinity of the sphere is $a^3 \, c_0$ so one
might be tempted to think naively that the free energy of depletion
could be something like $F_1\!\simeq\!a^3 \, c_0 \, k_{\rm B} T$.
But this disregards entirely the fact that the segments are all
connected (and are actually all on one single polymer chain because $a\!\ll\!\xi$).
A depleted test segment is connected to $h$ others where $a\!\simeq\!h^{3/5} \, A$,
in view of the excluded-volume effect. Therefore, the number of degrees
of freedom is reduced by a factor $h$
\begin{equation}
F_1 \simeq a^3 \, c_0 \, h^{-1} \, k_{\rm B} T \,,
\end{equation}
which agrees with Eq.(2). Note that this derivation is valid in the
mean for the effective number of degrees of freedom is actually less
than unity in Eq.(3).

In the following I discuss several problems concerning the depletion
interaction between nanospheres and polymer and its usefulness in
explaining the precipitation of proteins by polymer.

Water is often the solvent of choice in experiments concerning the
thermodynamic properties of polymer-nanocolloid or polymer-protein
mixtures. Although water-soluble polymers dissolve readily in
aqueous solution, the solvent must often be regarded as ``intermediate''
($\beta\!\ll\!A^3$) rather than ``excellent'' \cite{5}. The polymer chain then
interacts with a nanoparticle in a quasi-ideal manner \cite{12}.

In effect, the excluded-volume parameter $z\!=\!h^{1/2} \, \beta / A^3$
may remain smaller that unity if the nanoparticle is not too large.
In that case, a string of depleted segments behaves like a
Gaussian chain: $h\!\simeq\!a^2/A^2$ if the particle is a sphere.
This would imply the condition \cite{12} $a\!<\!A^4/\beta$ which may be easily
met in practice. Eq.(3) then leads to \cite{14}
\begin{equation}
F_1 = k_1 \, A^2 \, a \, c_0 \, k_{\rm B} T
= k_1 \, \left( \frac{a}{\xi_{\rm id}} \right) \, k_{\rm B} T \,,
\end{equation}
where $k_1$ is a numerical coefficient and $\xi_{\rm id} \!=\! 1 / (A^2 c_0)$
is a quasi-ideal correlation length. The latter relates to a state
in which the polymer chains are supposed hypothetically ideal.
Within the same approximation, one may argue in favor of a
self-consistent field picture at $a\!\ll\!\xi$ which leads to
a Laplace equation \cite{12} for $\Psi(\vec{r})$ where the inhomogeneous
polymer segment density $c(\vec{r})\!=\!c_0 \, \Psi(\vec{r})^2$
\begin{equation}
\Delta \Psi(\vec{r}) = 0 \,.
\end{equation}
The solution to Eq.(5) with $\Psi\!=\!0$ at the boundary of the sphere is
\begin{equation}
\Psi(r) = 1 - \frac{a}{r} \,.
\end{equation}
This leads to a value $k_1\!=\!2 \pi /3$ for the coefficient in Eq.(4).

Nevertheless, this SCF point of view cannot be entirely correct. At some
distance $r_{\ast}$ from the sphere, excluded volume effects must come into
play in Eq.(5) but then the argumentation for a mean-field approach also
becomes weak. From renormalization theory \cite{6}, we know that in the case
$\beta\!=\!A^3$ we have asymptotically
\begin{equation}
\Psi(r) - 1 \sim \left( \frac{a}{r} \right)^{\! \rm x} \,,
\end{equation}
with exponent x $\!\simeq\!4/3$.
When the solvent is intermediate ($\beta\!\ll\!A^3$), we again
adduce reasoning based on the parameter $z$ above to show that
Eq.(6) is only valid for $r\!<\!r_{\ast}\!=\!A^4/\beta$. Beyond $r_{\ast}$,
Eq.(6) must join smoothly to
\begin{equation}
\Psi(r) \simeq 1 - \frac{a \, A^{4/3}}{\beta^{1/3} \,\, r^{4/3}} \,.
\end{equation}
This reduces to Eq.(7) if $a\!>\!r_{\ast}$.

The SCF theory for the polymer distribution near a surface is incorrect
for a number of reasons. First, the correlation length $\xi$ is given
by a wrong power law which de Gennes \cite{2} proposed to amend by changing
the exponent in the excluded volume term in the SCF equation.
Nevertheless, the equation remains purely diffusive so it cannot
mimic the segment distribution about a sphere expressed by Eq.(7).
A further amendment could be to introduce a {\em fractional}
SCF equation. In the vicinity of the sphere, this would reduce to
\begin{equation}
r^{1-D} \frac{d}{dr} \left( r^{D-1} \, r^{-\Theta} \, \frac{d\Psi}{dr} \right) = 0 \,,
\end{equation}
which is a stationary generalized diffusion equation of fractal order $D$ and with
a diffusion coefficient $r^{-\Theta}$. (For a discussion of fractional diffusion
equations, see ref. \cite{15}). Eq.(7) imposes the constraint $D - \Theta \!=\! 10/3$
on the exponents $D$ and $\Theta$, which leaves one of them to be
suitably chosen. However, a fractional SCF equation is still not entirely
satisfactory. Density fluctuations are merely accounted for in a preaveraged sense.

It is interesting to note that there is a relation between the depleted
concentration about a sphere $c(r)-c_0$ and the pair correlation function
$g(r)$ pertaining to a single chain in the bulk \cite{13} (see Eqs.(6) and (7))
\begin{equation}
c(r) - c_0 \simeq - c_0 \, a^3 \, h^{-1} \, g(r) \,.
\end{equation}
This is independent of the strength of the excluded-volume effect.
The structure of the depletion hole is analogous to the pair correlation
structure of the polymer removed provided the latter is normalized by the
effective number of degrees of freedom depleted by the sphere.

It is of interest to consider the interaction between two nanospheres
separated at distance $R$ in the quasi-ideal limit. One needs to solve
Eq.(5) with boundary conditions $\Psi\!=\!0$ at their surfaces and
$\Psi\!=\!1$ at infinity. In the electrostatic analogy, the two spheres
are grounded. This is not a trivial problem \cite{16}. Using the method of images \cite{17},
one proceeds as follows. A positive test charge is placed at the center
of the first sphere. The potential $\Psi$ on the second sphere is brought
to zero by alternately adding appropriate image charges on the center
line between the two spheres: these are negative within the second sphere
but positive in the first. Next, a similar test charge of positive sign is
placed at the center of the second sphere. The potential on the first
sphere is now rendered uniform by again adding image charges on the
centerline but the signs are interchanged. The potential on both spheres
is identical and uniform. Thus, we derive the capacitance of the two
spheres as a series expansion which may be written in a less cumbersome
manner via the method of difference equations.
The depletion free energy turns out to be directly related to the capacitance
of a particle \cite{12} via Green's first identity which finally leads to
\begin{eqnarray}
\frac{F_2}{k_{\rm B} T} &=& \frac{A^2}{6} \, c_0 \, (C_2 - 8 \pi \, a) \,, \\
C_2 &=& 8 \pi \, a \, \sinh(\tau) \, \sum\limits_{n=1}^{\infty}
\frac{(-1)^{n+1}}{\sinh(n \, \tau)} \,, \\
\cosh(\tau) &\equiv& \frac{R}{2 a} \nonumber \,.
\end{eqnarray}
Eq.(11) disagrees with an expression quoted by Hanke {\em et al.} without
proof \cite{18}. The function $2 - C_2 / (4 \pi \, a)$ is very slowly varying
(see Table 1) and has a maximum equal to $2 - 2 \ln 2$ as the spheres
touch. Here, the units for capacitance have been chosen in such a way that
$C_1\!=\!4 \pi \, a$ for a single sphere. Inevitably, Eq.(11) breaks
down at separations $R$ beyond $r_{\ast}$ as the excluded-volume effect
starts to play a role. Nevertheless, the decay of $F_2$ with $R$ is not
fast enough for a finite computation of the second virial coefficient.
A cut-off at $R\!\simeq\!\xi$ must be introduced as has been discussed
by Eisenriegler \cite{19} in the excellent-solvent case ($\beta\!=\!A^3$).
The opposite limit of large spheres near ideal polymers has been addressed
by Tuinier {\em et al.} \cite{20}.

How well does an expression like Eq.(4) work? An important phenomenon
bioengineering is the precipitation of proteins by inert polymer.
This has been studied for a long time \cite{21,22,23,24,25} but the most thorough
quantitative study is that of Atha and Ingham \cite{26}. They determined the solubility $S$
of a host of proteins as a function of the polyethylene glycol (PEG)
added to the suspension. The concentrations of PEG were well into the
semidilute regime. The logarithm of the solubility turns out to be a
purely linear function of the PEG concentration and the resulting slopes
are a monotone function of the protein radius, the latter being an equivalent
quantity derived from the diffusion coefficient via the Stokes-Einstein
relation (see Fig. 1).
\begin{figure}
\centering
\includegraphics[angle=270,width=250pt]{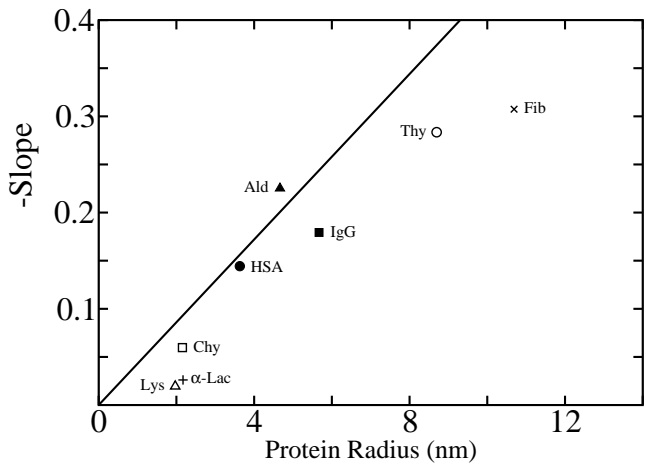}
\caption{The slope of the linear plot of $^{10\!}\log S$ versus $w$
as a function of the effective Stokes radius of various proteins.
The data are taken from ref. \cite{26}. The proteins are:
lysozyme (Lys), $\alpha$-lactalbumin ($\alpha$-Lac),
chymotrypsin (Chy), human serum albumin (HSA),
human $\gamma$-globulin (IgG), aldolase (Ald),
thyroglobulin (Thy), human fibrinogen (Fib)}
\end{figure}
The proteins may deviate significantly from an ideally spherical shape
so let us account for this fact. First, we know that for a compact
particle of general shape, Eq.(4) may be generalized to
\begin{equation}
F_1 = \frac{A^2}{6} \, C_1 \, c_0 \, k_{\rm B} T \,,
\end{equation}
in terms of the capacitance $C_1$ \cite{12}. Next, Hubbard and Douglas have shown
analytically and by simulation \cite{27,28} that the Brownian friction coefficient
of a particle of general shape is directly proportional to its capacitance
to an excellent approximation. (It is also well to recall that the
capacitance itself is essentially proportional the the particle's
surface area \cite{29,30}.) Since the chemical potential of a protein
given by the sum of $F_1$ and $k_{\rm B} T \, \ln S$ must be a
constant, experiment should conform to the expression
\begin{equation}
\frac{\Delta ^{10\!}\log S}{\Delta w} = -0.051 \, a_{\rm S} \,,
\end{equation}
valid for PEG solutions where $w$ is given in \% weight per unit
volume and $a_{\rm S}$ is the Stokes radius of the protein in nm.
(For data on PEG under theta conditions, see ref. \cite{31}.) In Fig. 1,
we have plotted Eq.(14) but with an adjusted coefficient {\bf -}0.043 $\!=\!$
{\bf -} (0.23/0.27) $\times$ 0.051 because PEG4000 isn't quite long enough
to be characterized as infinitely long (see Fig. 3 and Table 2 of ref. \cite{26}).
Although Eq.(14) overestimates the impact of polymer a bit, it is
remarkably how well a simple SCF model works. At high protein radii,
one expects significant upward deviations from linearity owing to
Eq.(2) but the trend is the reverse which is puzzling. Appreciable
attractive forces between a protein and PEG may sometimes exist as
has been suggested by Bloustine {\em et al.} \cite{32} (see also ref. \cite{33})
and could be the cause of these anomalies.

In this note and in previous work \cite{34}, we conclude that linear depletion
laws for thermodynamic quantities are valid up to quite high protein
concentrations. The linearity is in accord with scaling and SCF
arguments for nanoparticles. As yet there is little evidence for
interaction terms as given by Eq.(11) for example. It is quite
possible that repulsive forces between proteins (hard core, electrostatic)
are compensated by attractive forces (adhesive, depletion) in such
a way that quasi-ideal conditions apply. For a theory of this effect,
see ref. \cite{35}. The concentration of electrolyte added to the mixtures
in precipitation experiments \cite{26} would suggest that we are in
such a regime.

\vskip 20pt
\noindent
{\Large\bf Acknowledgment}
\vskip 10pt
\noindent
I thank Edgar Blokhuis for logistic help.


\begin{thebibliography}{99}

\bibitem{1} De Gennes, P.G. Rep. Prog. Phys. 1969, 32, 187.
\bibitem{2} De Gennes, P.G. Macromolecules 1981, 14, 1637.
\bibitem{3} De Gennes, P.G.; Pincus, P. J. Physique Lett. 1983, 44, 241.
\bibitem{4} De Gennes, P.G. C.R. Acad. Sci. Paris 1979, 288, 232.
\bibitem{5} Odijk, T Macromolecules 1996, 29, 1842.
\bibitem{6} Eisenriegler, E; Hanke, A; Dietrich, S. Phys. Rev. E 1996, 54, 1134.
\bibitem{7} Tuinier, R; Rieger, J; de Kruif, C.G. Adv. Coll. Int. Sci. 2003, 103, 1.
\bibitem{8} Eisenriegler, E. J. Chem. Phys. 2006, 125, 204903.
\bibitem{9} Eisenriegler, E. In ``Soft Matter, Volume 2'', eds. G. Gompper and M. Schick,
Wiley-VCH, 2005.
\bibitem{10} Hooper, J.B.; Schweizer, K.S.; Desai, T.G.; Koshy, R.;
Keblinski, P. J. Chem. Phys. 2004, 121, 6986.
\bibitem{11} Fuchs, M.; Schweizer, K.S. J. Phys. Cond. Mat. 2002, 14, R239.
\bibitem{12} Odijk, T. Physica A, 2000, 278, 347.
\bibitem{13} De Gennes, P.G. ``Scaling concepts in polymer physics'', Cornell University Press,
Ithaca, New York, 1979.
\bibitem{14} Odijk, T. Biophys. J. 2000, 79, 2314.
\bibitem{15} Hilfer, R. ``Applications of fractional calculus in Physics'', World Scientific,
Singapore, 2000.
\bibitem{16} Jeffrey, G.B. Proc. Roy. Soc. London A 1912, 87, 109.
\bibitem{17} Smythe, W.R. ``Static and dynamic electricity'', McGraw-Hill, New York, 1968.
\bibitem{18} Hanke, A.; Eisenriegler, E.; Dietrich, S. Phys. Rev. E 1999, 59, 6853.
\bibitem{19} Eisenriegler, E. J. Chem. Phys. 2000, 113, 5091.
\bibitem{20} Tuinier R.; Vliegenthart, G.A.; Lekkerkerker, H.N.W. J. Chem. Phys. 2000, 113, 10768.
\bibitem{21} Polson A.; Potgieter, G.M.; Largier, J.F.; Mears, G.E.F.; Joubert, F.J.
Biochim. Biophys. Acta 1964, 82, 463.
\bibitem{22} Juckes, I.R.M. Biochim. Biophys. Acta 1971, 82, 463.
\bibitem{23} H\"{o}nig, W.; Kula, M.R. Anal. Biochem. 1976, 72, 502.
\bibitem{24} Middaugh, C.R.; Lawson, E.O.; Litman, G.W.; Tisel, W.A.; Mood, D.A.; Rosenberg, A.
J. Biol. Chem. 1980, 255, 6532.
\bibitem{25} McPherson, A. ``Crystallisation of biological macromolecules'', Cold Spring Harbor
Laboratory Press, New York, 1999.
\bibitem{26} Atha, D.H.; Ingham, K.C. J. Biol. Chem. 1981, 256, 12108.
\bibitem{27} Hubbard, J.B.; Douglas, J.F. Phys. Rev. E 1993, 47, R2983.
\bibitem{28} Douglas, J.F.; Zhou, H.X.; Hubbard, J.B. Phys. Rev. E 1994, 49, 5319.
\bibitem{29} Russell, A. J. Inst. Elect. Eng. 1916, 55, 1.
\bibitem{30} Chow, Y.L.; Yovanovich, M.M. J. Appl. Phys. 1982, 53, 8470.
\bibitem{31} Kawagucki, S.; Imai, G.; Suzuki, J.; Mivahara, A.; Kitano, T.; Ito, K.
Polymer 1997, 38, 2885.
\bibitem{32} Bloustine, J.; Virmani, T.; Thurston, G.M.; Fraden, S. Phys. Rev. Lett. 2006, 96, 087803.
\bibitem{33} Sheth, S.R.; Leckband, D. Proc. Natl. Acad. Sci. USA 1997, 94, 8399.
\bibitem{34} Wang, S.; van Dijk, J.A.P.P.; Odijk, T.; Smit, J.A.M. Biomacromolecules 2001, 2, 1080.
\bibitem{35} Prinsen, P; Odijk, T. J. Chem. Phys. 2004, 121, 6525.

\end{thebibliography}
\end{document}